\documentclass[twocolumn,showpacs,preprintnumbers,amsmath,amssymb, prb, superscriptaddress]{revtex4-1}

\usepackage{graphicx}% Include figure files [pdftex]
\usepackage{dcolumn}% Align table columns on decimal point
\usepackage{bm}% bold math
\usepackage{latexsym}

\begin{document}

\preprint{APS/paper}

\title{Electric-field-induced monoclinic phase in (Ba,Sr)TiO$_3$ thin film}

\author{ A. S. Anokhin }
\affiliation{Southern Scientific Center RAS, 41, Chekhov str., Rostov-on-Don, 344006, Russia}
\affiliation{Faculty of Physics, Southern Federal University, 5, Zorge str., Rostov-on-Don, 344090, Russia}
\author{ Yu. I. Yuzyuk }
\affiliation{Faculty of Physics, Southern Federal University, 5, Zorge str., Rostov-on-Don, 344090, Russia}
\author{ Yu. I. Golovko }
\author{ V. M. Mukhortov }
\affiliation{Southern Scientific Center RAS, 41, Chekhov str., Rostov-on-Don, 344006, Russia}
\author{ M. El Marssi }
\affiliation{Universit$\acute{\mbox{e}}$ de Picardie Jules Verne, LPMC, 33 rue Saint-Leu,  F-80039 Amiens, France}

\date{\today}

\begin{abstract}
We have studied electric-field-induced symmetry lowering in the tetragonal (001)-oriented heteroepitaxial (Ba$_{0.8}$Sr$_{0.2}$)TiO$_3$ thin film deposited on (001)MgO substrate. Polarized micro-Raman spectra were recorded from the film area in between two planar electrodes deposited on the film surface. Presence of \textit{c}-domains with polarization normal to the substrate was confirmed from polarized Raman study under zero field, while splitting and hardening of the \textit{E}(TO) soft mode and polarization changes in the Raman spectra suggest monoclinic symmetry under external electric field.  
\end{abstract}


\maketitle

Ferroelectric materials exhibit spontaneously polarized phase below a critical temperature and their polarization can be switched between two symmetry equivalent directions under external electrical field.  Polarization switching in ferroelectric thin films is the functional basis of device applications \cite{c1,c2,c3} and their performance depends strongly on the magnitude and stability of the switchable polarization of the ferroelectric layer. In bulk materials, phase transition to the ferroelectric state is accompanied by the formation of domains to minimize the total energy of the system with respect to the depolarization field and mechanical strain. Additional boundary conditions in epitaxial films and superlattices change drastically their ferroelectric properties. The phase transition sequence and domain structure of ferroelectric thin films substantially differ from those of bulk ferroelectrics. The effects of internal stresses on the ferroelectric properties of thin films have been investigated theoretically and experimentally for a number of perovskites \cite{c4,c5,c6,c7,c8,c9,c10,c11,c12}. Ferroelectric properties of thin films are greatly influenced by epitaxial strain and polarization switching was found to be very sensitive to the nature of the epitaxial boundary conditions imposed by the substrate or/and film-electrode interface. Successful integration of ferroelectric thin films into microelectronic devices requires a better understanding of the influence of the external electric field on dielectric properties of epitaxial thin films.

External electric field changes the ionic positions and particular lattice vibrations, therefore, Raman spectroscopy is a powerful tool for thin films investigations under external field, however, up to date the effect of electric field on the lattice dynamics of ferroelectric thin films is not well studied so far. Electric-field-induced hardening of the soft modes in SrTiO$_3$ thin films using an indium-tin-oxide/SrTiO$_3$/SrRuO$_3$ heterostructure was reported by Akimov \textit{et al} \cite{c13}. The markedly different behavior of the soft modes in SrTiO$_3$ thin films from that in the bulk \cite{c14} was explained by the existence of local polar regions induced by oxygen vacancies. No influence on the Raman spectra under electric field bias was observed in (Ba$_{0.5}$Sr$_{0.5}$)TiO$_3$  thin films \cite{c15}. The applied field (36 kV/cm) was, probably insufficient to induce any change in the lattice dynamics.  In this letter, we report the electric-field-induced phase transition in heteroepitaxial (Ba$_{0.8}$Sr$_{0.2}$)TiO$_3$ thin film studied by micro-Raman spectroscopy.

Transparent and mirror-smooth (001)-oriented heteroepitaxial (Ba$_{0.8}$Sr$_{0.2}$)TiO$_3$ (BST) thin film (thickness 600 nm) was deposited on (001)MgO single crystalline 0.5 mm-thick substrate by rf sputtering of polycrystalline target of the corresponding stoichiometric composition. Details of the growth conditions have been previously reported \cite{c16}. The epitaxial relationships between the film and the MgO substrate have been confirmed by x-ray diffraction (XRD): $(001)_{\mbox{BST}} \parallel (001)_{\mbox{MgO}}$ and $[100]_{\mbox{BST}} \parallel [100]_{\mbox{MgO}}$. The lattice parameters normal to the substrate (\textit{c} = 0.4028 nm) and in plane of the film (\textit{a} = 0.3983 nm) were determined from the XRD patterns for (001)-oriented (\textit{c}-domain) film. At room temperature the tetragonality of the film was found larger (\textit{c/a} = 1.011) than that of the bulk ceramics of corresponding composition due to the two-dimensional clamping discussed previously \cite{c17,c18}. Al electrodes (800 nm-thick) were deposited on the top of the film by dc sputtering using standard optical photolithography. The electrodes width was 70 $\mu$m, while spacing between the electrodes was 2 $\mu$m only. The electrodes were aligned parallel to one of the substrate's cubic axes and prior to the Raman measurements the sample was cleaved along the direction perpendicular to the electrodes alignment in two parts with faces parallel to the cubic axes of the MgO substrate. Electrical connections were directly wire bounded to the electrodes to enable application of a dc voltage up to 600 kV/cm during Raman measurements.

\begin{figure}
\includegraphics{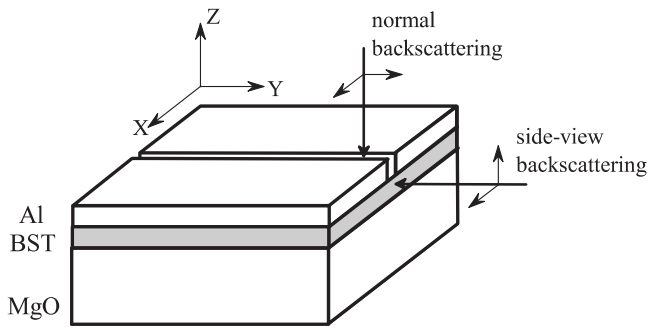}
\caption{\label{fig1} Side-view backscattering geometry used Raman scattering measurements (not to scale).}
\end{figure}

Raman spectra were excited using the polarized light of an argon ion laser ($\lambda$ = 514.5 nm) and analyzed using a Jobin Yvon T64000 spectrometer equipped with a charge coupled device. Polarized Raman spectra have been measured on the small sample carefully oriented according to the crystallographic axes of \textit{c}-domain film (X $\parallel$[100], Y $\parallel$[010] and Z $\parallel$[001]). An optical microscope with a 100x objective was used to focus on the sample the incident light as a spot of about 1 $\mu$m in diameter with the effective depth of focus about 2 $\mu$m. Polarized Raman spectra were recorded in normal and "side-view backscattering" geometries \cite{c17}. In the latter case the Raman microprobe beam was focused at the previously cleaved cross-section of the film area in between two electrodes (see Figure 1) that was viewed using a video microscope. The wave vector of the incident light was parallel to the substrate (Y axis) while polarization of the incident/scattered light was parallel or perpendicular to the \textit{c}-axis of the film (along X or Z axis) as shown in Fig. 1. 

\begin{figure}
\includegraphics{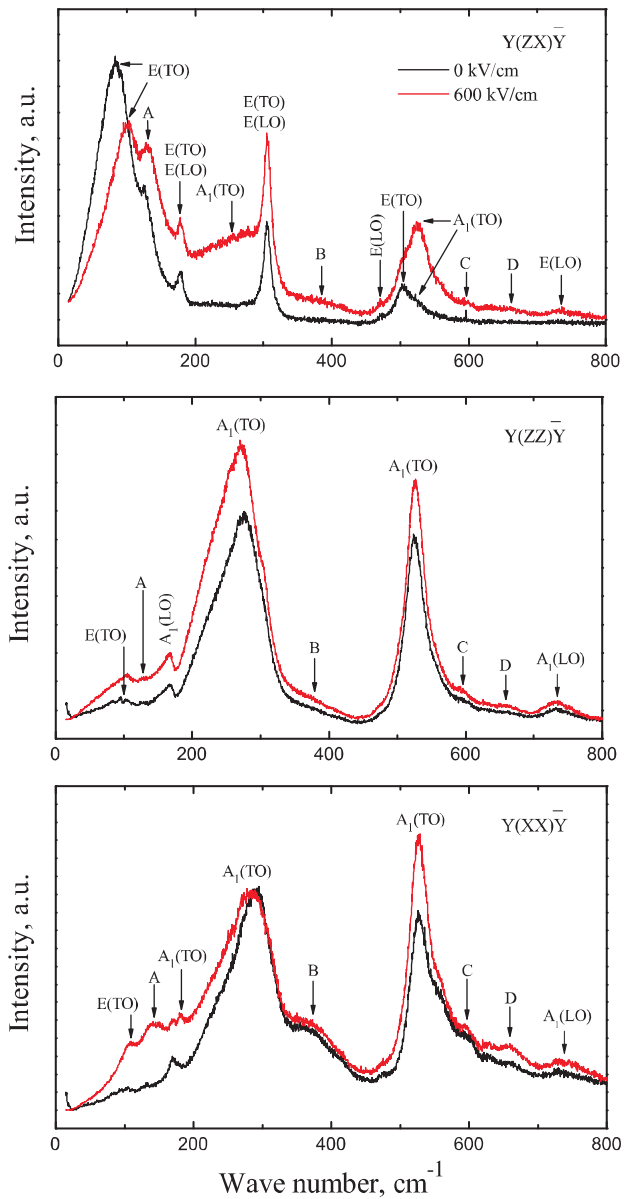}
\caption{\label{fig2} (Color on-line) Polarized Raman spectra of BST thin film recorded under zero field (black) and under external electric filed 600 kV/cm (red). All Raman spectra are corrected for the Bose-Einstein temperature factor. Mode assignments correspond to the tetragonal symmetry under zero field. All spectra presented in these scattering geometries correspond to $A^{\prime}$ modes of the monoclinic phase under external electric filed.}
\end{figure}

Polarized Raman spectra of BST thin film under zero field shown in Fig. 2, reflect all the typical features of BaTiO$_3$ (BT) single crystal and confirm the \textit{c}-domain orientation of the tetragonal film. Detailed mode assignment for similar BST thin film in the tetragonal ($C_{4V}$) ferroelectric phase based on the factor-group analysis was already reported \cite{c18}. The \textit{E} modes are allowed for $\alpha_{zx}$ and $\alpha_{zy}$ polarizability tensor components and, therefore, can be obtained only in a "side-view backscattering" geometry. Figure 2 shows $A_1$ and \textit{E} phonons in the tetragonal c-domain phase. Certainly, the weak 260 and 523 cm$^{-1}$ bands appear in the   spectrum due to the leakage of the $A_1$(TO) intense modes. In turn, the partial leakage of the \textit{E}(TO) soft mode into the diagonal geometries occurs. It is worth mention that similar leakage was also observed even in single-domain BT single crystals \cite{c19,c20}. There are additional disorder-activated bands centered at ~129, 390, 570, and 640 cm$^{-1}$ (marked as \textit{A, B, C,} and \textit{D}, respectively, in Fig. 2), their origin due to the disorder of Ti ions was already discussed\cite{c18}. Most of the peaks in the Raman spectra of BST thin film under zero field are only slightly shifted with respect to their analogs in BT, whereas the soft mode components are markedly altered. As shown in Fig. 2, the \textit{E}(TO) soft mode appears as \textit{underdamped peak} centered at 84 cm$^{-1}$ with the half width of 62 cm$^{-1}$, while in bulk ceramics of BT and Ba-rich (Ba content more than $70\%$) BST this mode is overdamped with the damped harmonic oscillator fitted wave number being 35$\pm$5 cm$^{-1}$ and the half width larger than 100 cm$^{-1}$. In the \textit{c}-domain thin film, the \textit{E}(TO) soft mode corresponds to Ti displacement with respect to oxygen octahedra in the \textit{xy} plane \cite{c21}, the plane being parallel to the substrate. Therefore this mode is very sensitive to the two-dimensional (2D) clamping imposed on the film by the substrate. On cooling from the growth temperature ($\sim$1000 K) 2D stresses develop gradually due to the difference in thermal-expansion coefficients of the film and substrate and induces enhanced tetragonality in the film. According to our previous investigations 2D stress vanishes in partially unsupported and completely free-standing BST thin films \cite{c22}. Polarized Raman spectra at the film area far away from the top electrodes were also examined and no differences in the peak positions were found implying that deposition of electrodes does not affect the vibrational spectra of the film.      

\begin{figure}
\includegraphics{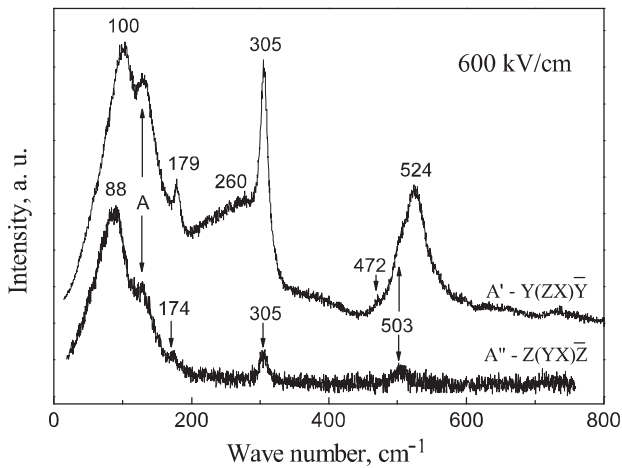}
\caption{\label{fig3} Polarized Raman spectra of BST thin film recorded in crossed scattering geometries under external electric filed 600 kV/cm. All Raman spectra are corrected for the Bose-Einstein temperature factor. }
\end{figure}

According to the order-disorder model \cite{c23}, in the paraelectric phase of BT the Ti ion does not have a potential energy minimum at the center of the unit cell, but has eight minima along the [111] pseudocubic axes. In the tetragonal phase under zero external field, four sites along a face of the unit cell become energetically favored, with a time-averaged position along the \textit{c} axis. Under electric field applied perpendicular to the polar axis Ti ions preferentially occupy only two of the eight sites as a result the perovskite primitive cell should exhibit monoclinic distortion.

According to the Curie principle, external electric field (group symmetry $C_{\infty \mbox{v}}$) applied in the direction perpendicular to the polar axis of the tetragonal BST film with the symmetry $C_{4\mbox{v}}$ ($C_{4}\parallel O_{Z}$) induces the symmetry lowering so that resulting group symmetry is monoclinic - $C_{s}$ ($\sigma_{xz}$). Corresponding Raman selection rules are presented in Table I. Partial depolarization observed in the Raman spectra recorded under external field 600 kV/cm (Fig. 2) applied in the direction perpendicular to the spontaneous polarization of the \textit{c}-domain film (i.e. along the X axis) is in agreement with the monoclinic symmetry. Namely, intensity of the $A_1$(TO) modes at 260 and 523 cm$^{-1}$ increases considerably in the Y(ZX)$\overline{\mbox{Y}}$ spectrum, while components of the \textit{E}(TO) soft mode appears in the Y(XX)$\overline{\mbox{Y}}$ and Y(ZZ)$\overline{\mbox{Y}}$ spectra. Figure 3 shows Raman spectra under external field 600 kV/cm in the two crossed polarized scattering geometries where splitting of the \textit{E} soft mode into $A^{\prime}$(Y(ZX)$\overline{\mbox{Y}}$) component at 100 cm$^{-1}$ and $A^{\prime\prime}$(Z(YX)$\overline{\mbox{Z}}$) component at 88 cm$^{-1}$ is seen clearly. The activation of the $A_1$(TO) modes in the Y(ZX)$\overline{\mbox{Y}}$ crossed polarized spectrum and their absence in the Z(YX)$\overline{\mbox{Z}}$ spectrum correlate with the selection rules presented in Table I and confirm electric-field-induced phase transition to the monoclinic phase. All modes other than soft modes do not exhibit remarkable electric-field induced shits. Both $A^{\prime}$ and $A^{\prime\prime}$ components originating from the \textit{E}(TO) soft mode exhibit upward shift while $A^{\prime}$ component originating from $A_1$(TO) component of the soft exhibits downward shift implying decreasing of the tetragonal distortion due to polarization rotation within the monoclinic plane. Deviation of the polarization from the direction normal to the substrate depends on external field applied. The high-frequency modes (472 and 503 cm$^{-1}$), originating from \textit{E} modes of the tetragonal phase exhibit rather weak, or even no splitting into $A^{\prime}$ and $A^{\prime\prime}$ components in the monoclinic phase.

\begin{figure}
\includegraphics{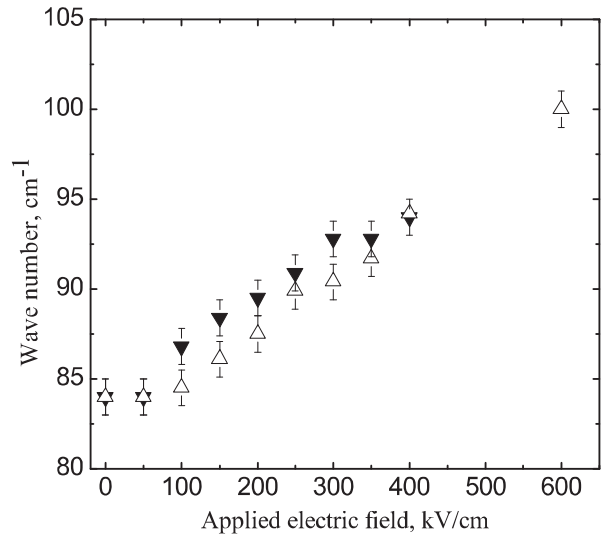}
\caption{\label{fig4} Wave number of the soft mode as a function of external dc electric field. Open symbols – increasing of the applied voltage, full symbols – decreasing.}
\end{figure}

In the tetragonal phase the $A_1$ and \textit{E} modes are associated with the clamped dielectric constant along and perpendicular to the polar \textit{c}-axis, respectively. Therefore, observed decreasing of the $A_1$ ($A^{\prime}$) - \textit{E} ($A^{\prime}$ + $A^{\prime\prime}$) splitting in the electric-field-induced monoclinic phase suggests increasing of the dielectric constant in the direction perpendicular to the film surface and decreasing in the film plane. The $A^{\prime}$ component of the former \textit{E}(TO) soft mode exhibits gradual hardening (see Fig. 4) under external electric field up to 100 cm$^{-1}$ at 600 kV/cm. Note that electric field effect is seen in the Raman spectra only above 100 kV/cm. No significant hysteresis of the soft mode wave number was observed during increasing-decreasing applied voltage cycles. After voltage application, all polarized Raman spectra were recorded again at zero field and no differences with respect to virgin sample were found.

In conclusion, we demonstrated electric-field-induced symmetry lowering in the heteroepitaxial BST thin film.  In the polarized Raman spectra recorded under electrical field applied in the direction perpendicular to the spontaneous polarization of the \textit{c}-domain film we observed remarkable shift of the soft mode and polarization changes in the Raman spectra which give an evidence of the monoclinic distortion.  

\begin{table}
\caption{\label{tab1}
Raman selection rules for tetragonal and monoclinic point groups.}
\begin{ruledtabular}
\begin{tabular}{ccc}
Raman activity& $ C_{4v} \rightarrow C_{s} $ &  Raman activity \\ \hline \\
XX, YY, ZZ     &  $A_{1} \rightarrow  A^{\prime}$   &  XX, YY, ZZ, XZ, ZX \\
XX, YY    &  $B_{1} \rightarrow  A^{\prime}$   & XX, YY, ZZ, XZ, ZX \\
XZ, ZX &  $E \rightarrow  A^{\prime}$   &XX, YY, ZZ, XZ, ZX\\
XZ, ZX &  $E \rightarrow  A^{\prime\prime}$   &XY, YX, YZ, ZY\\
\end{tabular}
\end{ruledtabular}
\end{table}

\begin{acknowledgments}
This study was partially supported by the Russian Foundation for Basic Research (RFBR), Grant Nos. $09-02-00666\_a$ and $09-02-00254\_a$. YY thanks UPJV for financial support.
\end{acknowledgments}

%\bibliography{paper}

\end{document}